\documentclass[preprint2]{aastex}
\usepackage{natbib}
\slugcomment{to appear January 2005 edition of AJ}

\shorttitle{A possible third component in DENIS-P~J020529.0-115925}
\shortauthors{Bouy, Mart\'\i n, Brandner \& Bouvier}

\begin{document}


\title{A possible third component in the L dwarf binary system DENIS-P~J020529.0-115925 discovered with the Hubble Space Telescope}


\author{H. Bouy }

\affil{Max-Planck  Institut f\"ur extraterrestrische Physik, Giessenbachstra\ss e 1,  D-85748 Garching bei M\"unchen, Germany}
\affil{Laboratoire d'Astrophysique de l'Observatoire de Grenoble, 414 rue de la piscine, F-38400 Saint Martin d'H\`ere, France}
\email{hbouy@mpe.mpg.de}

\and 

\author{E.~L. Mart\'\i n }
\affil{Instituto de Astrofisica de Canarias, 38200 La Laguna, Spain}
\affil{University of Central Florida, Department of Physics, PO Box 162385, Orlando, FL 32816-2385, USA}
\email{ege@iac.es}

\and

\author{W. Brandner}
\affil{Max-Planck Institut f\"ur Astronomie, K\"onigstuhl 17, D-69117 Heidelberg, Germany}
\email{brandner@mpia.de}

\and

\author{J. Bouvier}
\affil{Laboratoire d'Astrophysique de l'Observatoire de Grenoble, 414 rue de la piscine, F-38400 Saint Martin d'H\`ere, France}
\email{Jerome.Bouvier@obs.ujf-grenoble.fr}




\begin{abstract}
We present results showing that the multiple system DENIS-P~J020529.0-115925 is likely to be a triple system of brown dwarfs. The secondary of this previously known binary system shows a clear elongation on six images obtained at six different epochs. Significant residuals remain after PSF subtraction on these images, characteristic of multiplicity, and indicating that the secondary is probably a double itself. Dual-PSF fitting shows that the shape of the secondary is consistent with that of a binary system. These measurements show that the probability that DENIS-P~J020529.0-115925 is a triple system is very high. The photometric and spectroscopic properties of the system are consistent with the presence of three components with spectral types L5, L8 and T0.
\end{abstract}





\section{Introduction}
Multiple systems are important tests for the models of formation and evolution of very low mass stars and brown dwarfs. The binary fraction reported recently by \citet{2003AJ....126.1526B, 2003ApJ...586..512B,2003ApJ...587..407C, 2003AJ....125.3302G,2003ApJ...594..525M} for very low mass stars and brown dwarfs (between 10$\sim$15\%) cannot be well reproduced by the so-called ``photo-evaporation'' and ``embryo-ejection models'', which predict a much lower binary frequency \citep[$<$5\%; see e.g][ for a discussion on the different scenarii of formation mentioned here]{2003MNRAS.346..369K}. On the other hand the ``star-like'' model cannot explain the observed distribution of separation (all less than 20~A.U, with a peak around 4$\sim$8~A.U), and of mass ratios (with a strong lack of multiple systems with large mass ratios). The overall process of formation of brown dwarfs is still not well understood. The existence itself of an old triple system made of brown dwarfs would give extremely important and new constrains on these models, since no such object has been observed yet. 

In section \ref{denis0205}, we will present a summary of the known properties of DENIS-P~J020529.0-115925. In section \ref{obs} we will describe the observations. In section \ref{data_analysis}, we will present the analysis of the data leading to the discovery of a possible third component, and in section \ref{discussion} we will discuss about the triple system and its properties.

\section{DENIS-P~J020529.0-115925 \label{denis0205}}
DENIS-P~J020529.0-115925 is a L5 field very low mass dwarf \citep{1999AJ....118.2466M}. From their photometry and spectroscopy, \citet{2002ApJ...564..466G} classify the object as a L5.5$\pm$2 dwarf. It is classified as L7 in the \citet{2000AJ....120..447K} scheme. We adopt a spectral class of L5. It was discovered by \citet{1997A&A...327L..25D}  and first resolved as a binary by \citet{1999ApJ...526L..25K} using Keck images. This object has been well studied and observed, and is reported in several surveys (DENIS, and 2MASS, as \object{2MASSW~J0205293-115930}). Several authors \citep{1997A&A...327L..25D, 2001ApJ...561L.115M, 2003ApJ...586..512B} report methane absorption in their spectra, which implies a mass below the sub-stellar limit and an effective temperature less than 1800~K, as stated by \citet{2002ApJ...566..435S}. \citet{1999AJ....117.1010T} report similar absorption features in their  spectra but attribute it to H$_{2}$ rather than CH$_{4}$. \citet{1997A&A...327L..29M} reported a non-detection of lithium absorption from high resolution optical spectra, and inferred a lower limit on the mass of DENIS-P~J020529.0-115925~A of 60 Jupiter masses. Its distance (19.76$\pm$0.57~pc) and proper motion (437.8$\pm$0.8~mas/yr with P.A=82.8$\pm$0.1\degr) have been measured via trigonometric parallax \citep{2002AJ....124.1170D}. \citet{2000ApJ...538..363B} also measured its rotational velocity (22$\pm$5~km/s), and from the Cs~I and Rb~I absorption, they estimated its temperature to be T$_{eff}$=1700$\sim$1800~K.

\section{Observations \label{obs}}
We observed DENIS-P~J020529.0-115925 using the Hubble Space Telescope (HST) at six different epochs. The observations occurred between October 2000 and December 2003 during HST Cycles 8, 9, 10, 11 and 12 (program GO8720, P.I. Brandner, and programs GO9157, GO9345 and GO9968, P.I. Mart\'\i n). DENIS-P~J020529.0-115925 was observed with the Planetary Camera of WFPC2 \citep{WFPC..Instrument..Handbook}, in the F675W and F814W filters (GO8720) and F606W and F814W filters (GO9157, GO9345 and GO9968). Table \ref{obs_log} gives a log of all the observations we used for this study. The target is very red, and the observations in the F814W filter were more sensitive than the one in the F675W or F606W filters, despite the shorter exposure times and the lower quantum efficiency of WFPC2 at longer wavelengths. The possible third component does not appear clearly on the F675W and F606W images. For these reasons, the following analysis is made essentially in the F814W filter. The high angular resolution of the WFPC2-PC (0\farcs0455 pixel scale) allowed us to resolve easily the 0\farcs3 secondary, and to discover that it is elongated.

\section{Data analysis \label{data_analysis}}

We processed the data in two steps: a point spread function (PSF) subtraction on each component of the binary to look for residuals indicating the presence of a third companion, and a dual-PSF fitting of the secondary for confirmation.

\subsection{PSF Subtraction and Residuals}

\begin{figure*}
   \centering
   \includegraphics[width=\textwidth]{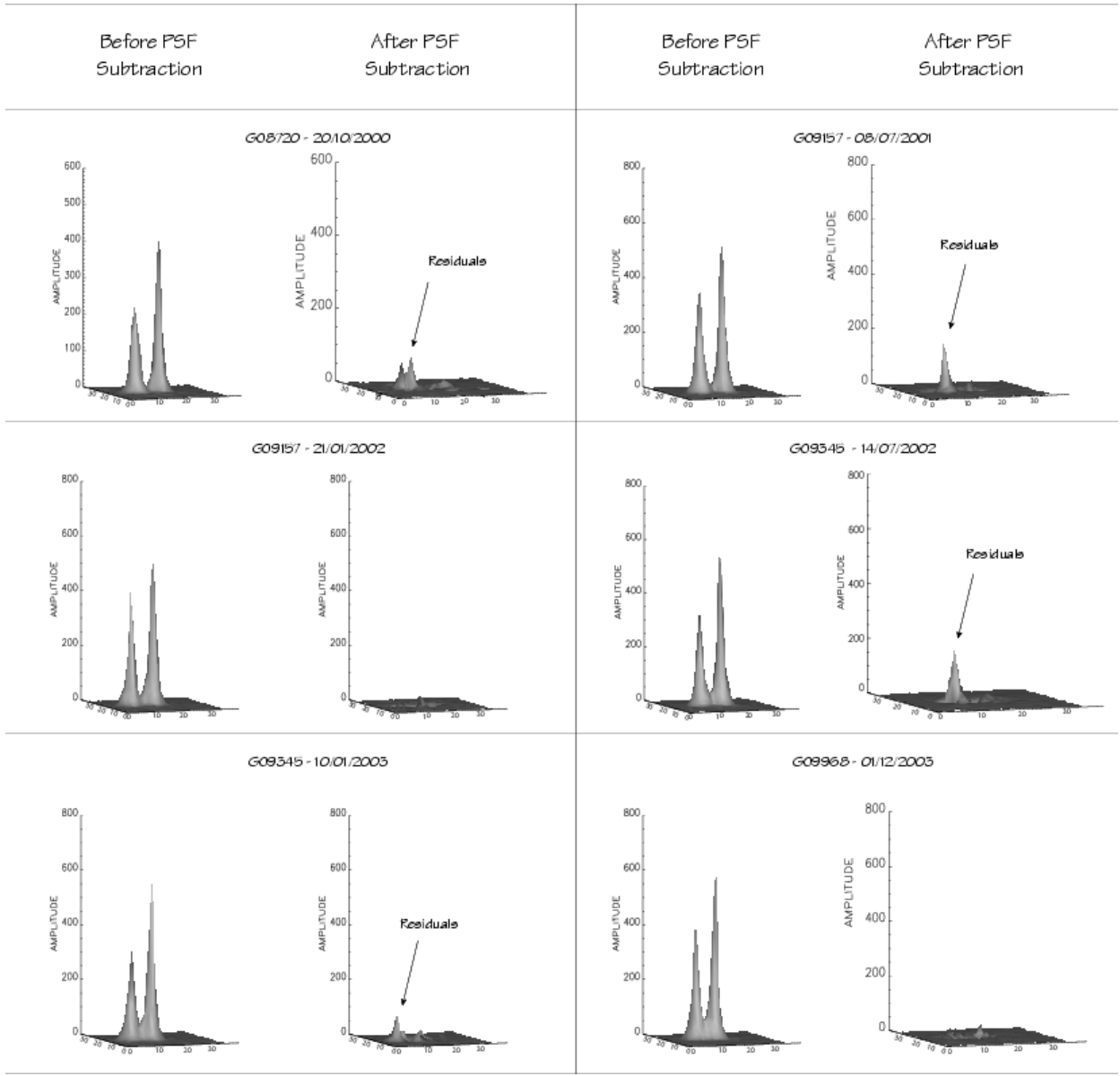}
     \caption{Results of the PSF subtraction at different epochs. This figure shows surface plots of the results of the PSF subtraction with one of the 10 reference PSF used. The primary is well subtracted whereas much stronger residuals remain for the secondary (clearly noticeable in 4 of the 6 epochs, indicated by arrows), indicating the presence of a third component. Similar results are obtained with the 9 other reference PSF stars. X and Y axis are in pixels, Z axis is in counts.\label{psf_subtraction}}
   \end{figure*}

Taking advantage of the extremely stable PSF of HST/WFPC2, we performed a PSF subtraction on the six images of DENIS-P~J020529.0-115925 using 9 different unresolved objects from programs GO8720 \citep[P.I. Brandner,][]{2003AJ....126.1526B} and GO8581 \citep[P.I. Reid,][]{2003AJ....125.3302G}, and 1 synthetic PSF obtained with \emph{Tiny Tim} \citep{TINY_TIM} as reference PSF stars. Figure \ref{psf_subtraction} shows that on 4 of the 6 images, the primary is extremely well subtracted, while strong residuals appears clearly after subtraction of the secondary, indicating the presence of a third component. The results with the 9 other reference PSF stars are similar to that presented in Figure \ref{psf_subtraction}. 

Using this technique, we tried to define a criterion that would allow to determine quantitatively if the secondary is a multiple system or not. We performed a detailed PSF analysis of all the WFPC2 images of programs GO8581 \citep[P.I. Reid][]{2003AJ....125.3302G}. This large sample has the advantage to be homogeneous (all targets were observed with the F814W filter of the WFPC2/PC), and all targets are very low mass stars and brown dwarfs with spectral properties (and therefore PSF) similar to DENIS-P~J020529.0-115925. 

The relative intensity of the residuals, defined as the integrated intensity of the residuals after PSF subtraction, divided by the integrated intensity of the object as expressed in equation \ref{eq:eq1}, appears to be a powerful criterion to identify binary candidates.
\begin{equation}
\mathcal{R.I}=\frac{\displaystyle \sum_{i=1}^{n} \mathcal{F}_{R}^{2}(i)}{\displaystyle \sum_{i=1}^{n} \mathcal{F}_{O}^{2}(i)} \label{eq:eq1}
\end{equation}
where: $n$ is the number of pixels, $\mathcal{F}_{R}(i)$ is the flux in pixel $i$ after PSF subtraction, and $\mathcal{F}_{O}(i)$ is the flux in pixel $i$ before PSF subtraction. 

\begin{figure*}
   \centering
   \includegraphics[width=\textwidth]{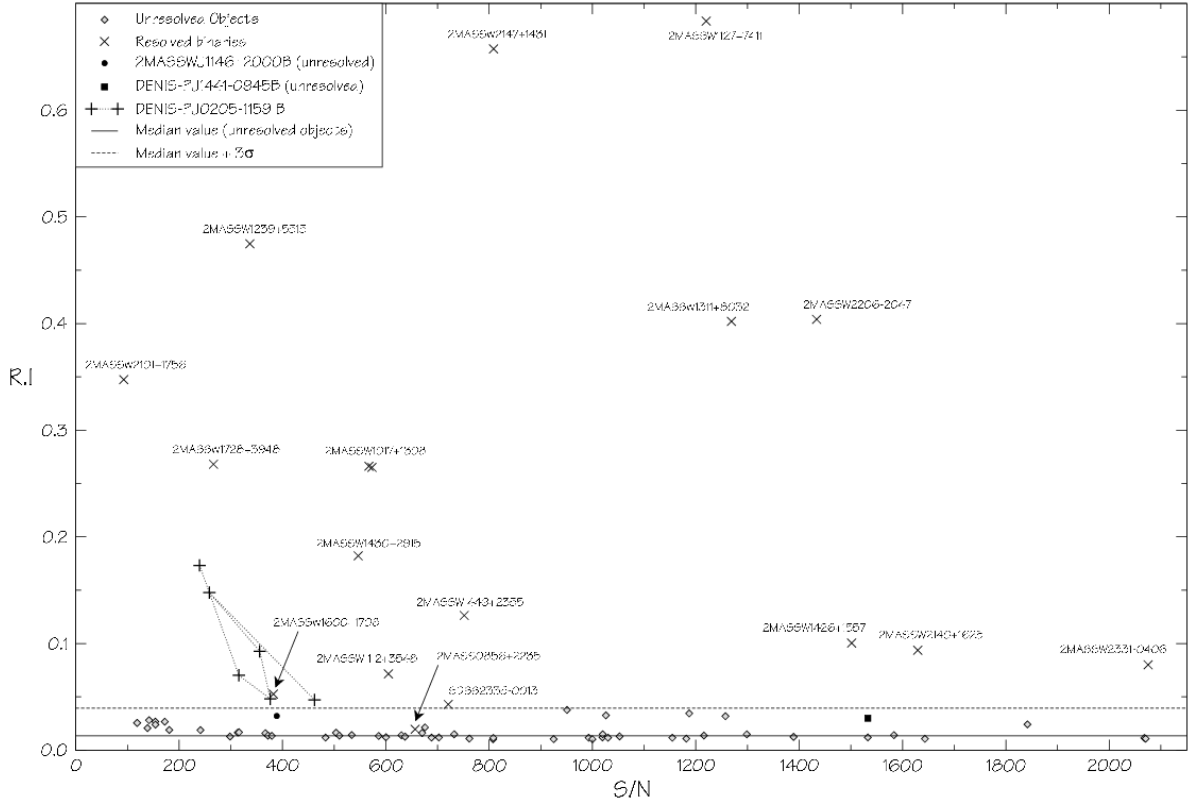}
      \caption{Residuals after PSF subtraction on HST/WFPC2 images (F814W filter) of very low mass stars and brown dwarfs. This plot shows $\mathcal{R.I}$ as a function of the signal-to-noise ratio for the targets of program GO8581. The crosses represent the multiple systems. The plus represent the values obtained at 6 epochs for DENIS-P~J020529.0-115925B. A filled square and a disk represent the results for two unresolved companions of known multiple systems, respectively 2MASSW~J1146+2000B (program GO8146, P.I. Reid) and DENIS-P~J1441-0945B (program GO8720, P.I. Brandner). The 3-$\sigma$ above the median and the median values for the unresolved objects are represented. This figure shows that the intensity of the residuals after PSF subtraction appears to be a powerful indicator of multiplicity. DENIS-P~J020529.0-115925B appears clearly different from the unresolved objects and from the unresolved companions of known multiple systems, while its residuals are comparable to that of known close multiple systems.}
      \label{residuals}
   \end{figure*}

Figure \ref{residuals} shows that $\mathcal{R.I}$ is very low and very stable for unresolved objects, while it is always higher than $\sim$3-$\sigma$ above the median value in the case of multiple systems. This technique found easily and automatically all the multiple systems present in the sample \citep[see ][]{2003AJ....126.1526B,2003AJ....125.3302G}, except one (2MASSW~J0856479+223518). This latter multiple system is indeed a good example to illustrate the limitations of this technique. 2MASSW~J0856479+223518 is a very close binary ($\delta$=0\farcs1) with a relatively large difference of magnitude \citep[$\Delta$Mag=2.8~mag, see ][]{2003AJ....126.1526B}. The relative intensity of the residuals after PSF subtraction of the primary, corresponding roughly to the ratio of the intensity of the secondary over that of the primary, is therefore of the order of a few percent only ($\Delta$Mag=2.8~mag is equivalent to a flux ratio of $f_{B}/f_{A}\sim$0.07). The relative intensity of the residuals after PSF subtraction is therefore a good method to find multiple systems with moderate differences of magnitude. Fortunately, it is the case of the possible third component of DENIS-P~J020529.0-115925, as shown below.

Using this property, we compare the relative intensity of the residuals after PSF subtraction on the secondary of DENIS-P~J020529.0-115925, using the same library of PSF stars. Figure \ref{residuals} shows the results. The residuals at six different epochs are all higher than that of unresolved objects and unresolved companions of known binaries of the same sample, but very similar to that of other multiple systems. This figure shows clearly that DENIS-P~J020529.0-115925B is very likely to be a binary system itself. Figure \ref{residuals} shows also that even for the two images where the residuals do not appear clearly by eyes (see Figure \ref{psf_subtraction}), the $\mathcal{R.I}$ is slightly higher than in any unresolved object. 

One should note that $\mathcal{R.I}$ strongly depends on the signal-to-noise ratio (S/N). At low S/N, the value of $\mathcal{R.I}$ must indeed increase due to the increased noise in the target. Figure \ref{residuals} shows that all our images were obtained at sufficiently high S/N (always greater than 100), so that $\mathcal{R.I}$ provides a robust statistical test of multiplicity.

\subsection{PSF fitting}

\begin{figure*}
   \centering
   \includegraphics[height=0.9\textheight]{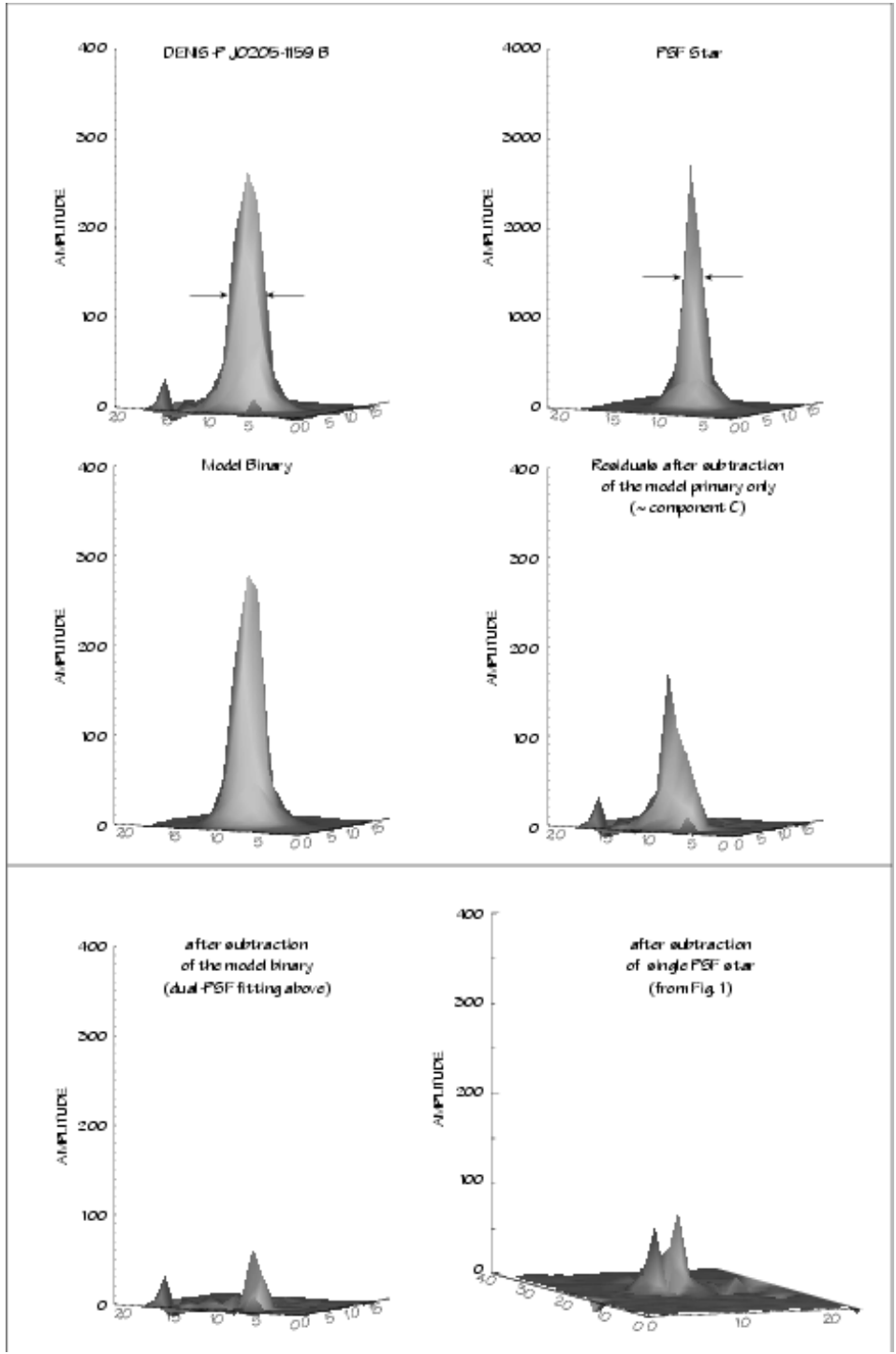}
      \caption{Results of the dual-PSF fitting on DENIS-P~J020529.0-115925B for the GO8720 image. As shown with arrows, DENIS-P~J020529.0-115925B has a FWHM much wider than the reference PSF star. The companion appears clearly after subtraction of the model primary. The residuals are small (lower panel, on the left), indicating that the quality of the fit is good. They are especially smaller than that obtained with a single PSF subtraction (lower panel, on the right). Similar results have been obtained with the 9 other reference PSF stars, showing that the result of the fit is robust and reliable.  X and Y axis are in pixels, Z axis is in counts.}
      \label{companion}
   \end{figure*}

In order to confirm that these residuals are consistent with the presence of a third component, we performed a dual-PSF fitting, on the secondary only, using a custom-made program described in \citet{2003AJ....126.1526B}. Briefly, and as presented in Figure \ref{companion}, the PSF fitting routine builds a model binary using each of the ten different PSF stars mentioned above, and then performs a non-linear PSF fit of the observed image to determine the best-fit values for the three free parameters: separation, position angle and flux ratio. The average of the ten results gives the final values for the parameters. Figure \ref{companion} shows that the residuals after this dual-fit are much better than after the single-PSF subtraction described in the previous section and in Figure \ref{psf_subtraction}. Moreover, the stability of the results for the ten different PSF stars at each of the six epochs shows that the results are reliable and robust, and that the third component hypothesis is consistent with the shape of the elongated PSF and highly probable.

Table \ref{flux_ratio} gives the flux ratios between the three components of the multiple system estimated with this method, as well as the corresponding differences of magnitude. According to the most recent DUSTY models of \citet{2000ApJ...542..464C}, these differences of magnitudes indicate that the three components must have similar masses. A difference of magnitude of 1.5~mag in the F814W corresponds indeed to a mass ratio of 75, 85 and 95\% at respectively 0.5, 1.0, and 5~Gyr. 

\section{Discussion \label{discussion}}

\subsection{Properties of the triple system}
The present analysis of the high angular resolution images indicates that DENIS-P~J020529.0-115925 is very likely to be a triple system. 

\begin{figure*}
   \centering
   \includegraphics[width=\textwidth]{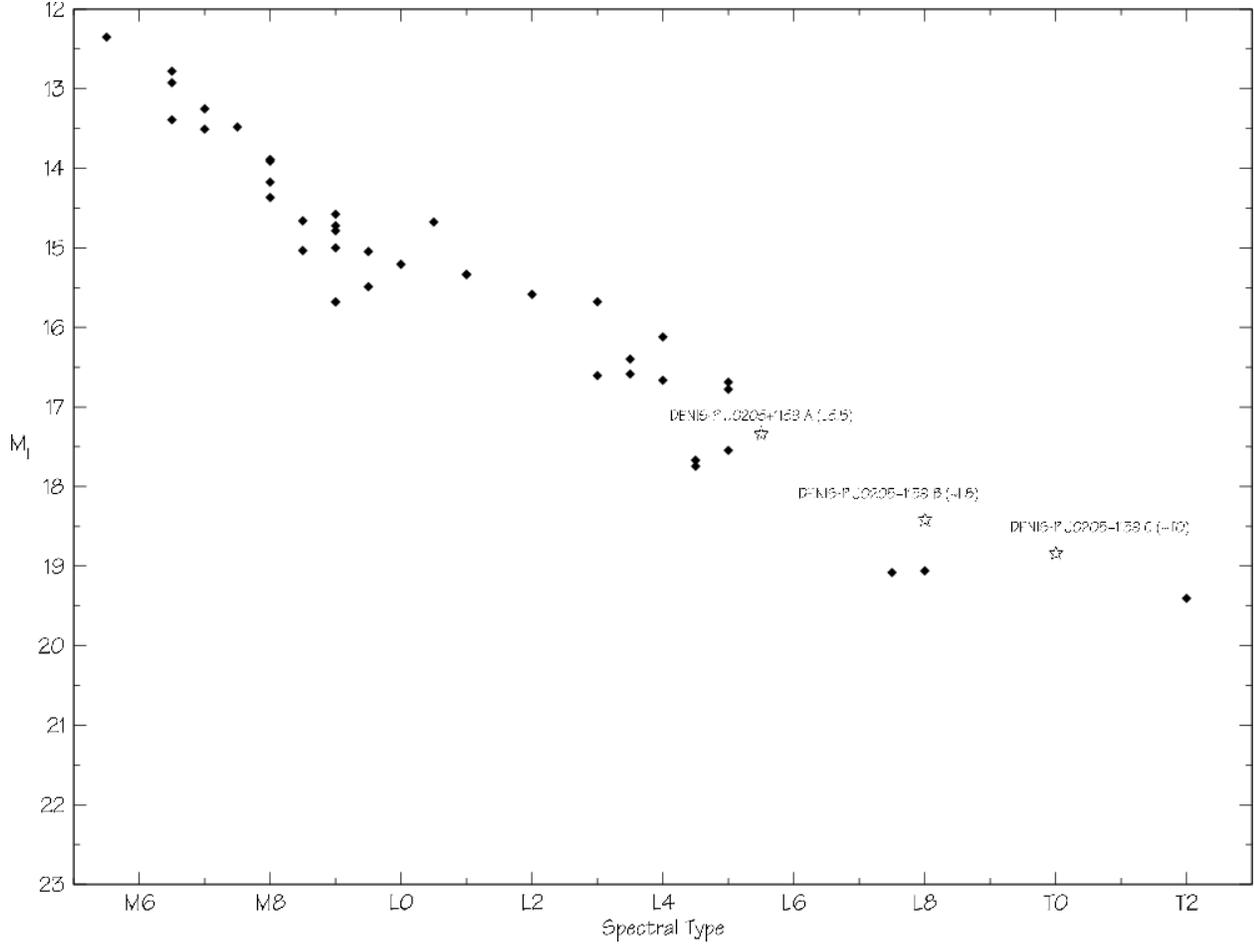}
      \caption{M$_{\mathrm{I_{C}}}$ vs Spectral Type relation. All measurements from \citet{2002AJ....124.1170D} (See their Figure 2, and Tables 1, 2 and 3) except for DENIS-P~J020529.0-115925A, B and C, evaluated as explained in the text. The absolute magnitudes of DENIS-P~J020529.0-115925A, B and C have been estimated using the differences of magnitude reported in Table \ref{flux_ratio}, and the DENIS I magnitude of the unresolved system.}
      \label{Mi_spt}
   \end{figure*}

Figure \ref{Mi_spt} shows the M$_{\mathrm{I_{C}}}$ vs Spectral Type relation for all the objects reported by \citet{2002AJ....124.1170D} (see their Tables 1, 2 and 3). Assuming differences of magnitude in I$_{\mathrm{C}}$ equal to that reported in Table \ref{flux_ratio} for the F814W filter \citep[F814W filter is close to I$_{\mathrm{C}}$,][]{WFPC..Instrument..Handbook}, and the DENIS I magnitude\footnote{ I$_{\mathrm{DENIS}}$ is very close to the I$_{\mathrm{Cousins}}$ \citep{Delfosse_PhD}} of the unresolved objects, one can estimate the spectral types of the three components. DENIS-P~J020529.0-115925~A is consistent with a L5.5 dwarf, in good agreement with the measurement reported by \citet{1999AJ....118.2466M} and \citet{2002ApJ...564..466G}, and showing that the primary would be dominating the optical spectrum. B and C would be consistent with $\sim$L8 and $\sim$T0 dwarfs respectively, in the \citet{2002ApJ...564..466G} classification scheme.

Several authors report the detection of methane absorption in the infrared spectrum of DENIS-P~J020529.0-115925 \citep{1997A&A...327L..25D, 2001ApJ...561L.115M, 2003ApJ...586..512B}. \citet{1999AJ....117.1010T} also observed this absorption feature in the spectrum but attribute it to H$_{2}$ rather than methane. \citet{2002ApJ...564..421B} note that this feature is weak and variable, and consider that it does not constitute a clear detection of methane. We note that if real, this feature could be related to the presence of L8 and T0 companions, and its variability to some weather effects, as already observed in other late L and T dwarfs by \citet{2003AJ....126.1006E}. \citet{2002AJ....124.1170D} showed that the absolute J and K-band magnitudes of early T-dwarfs are similar to that of late-L dwarfs, so that the contribution of B and C in the near-infrared can be larger than in the optical.

\begin{figure*}
   \centering
   \includegraphics[width=\textwidth]{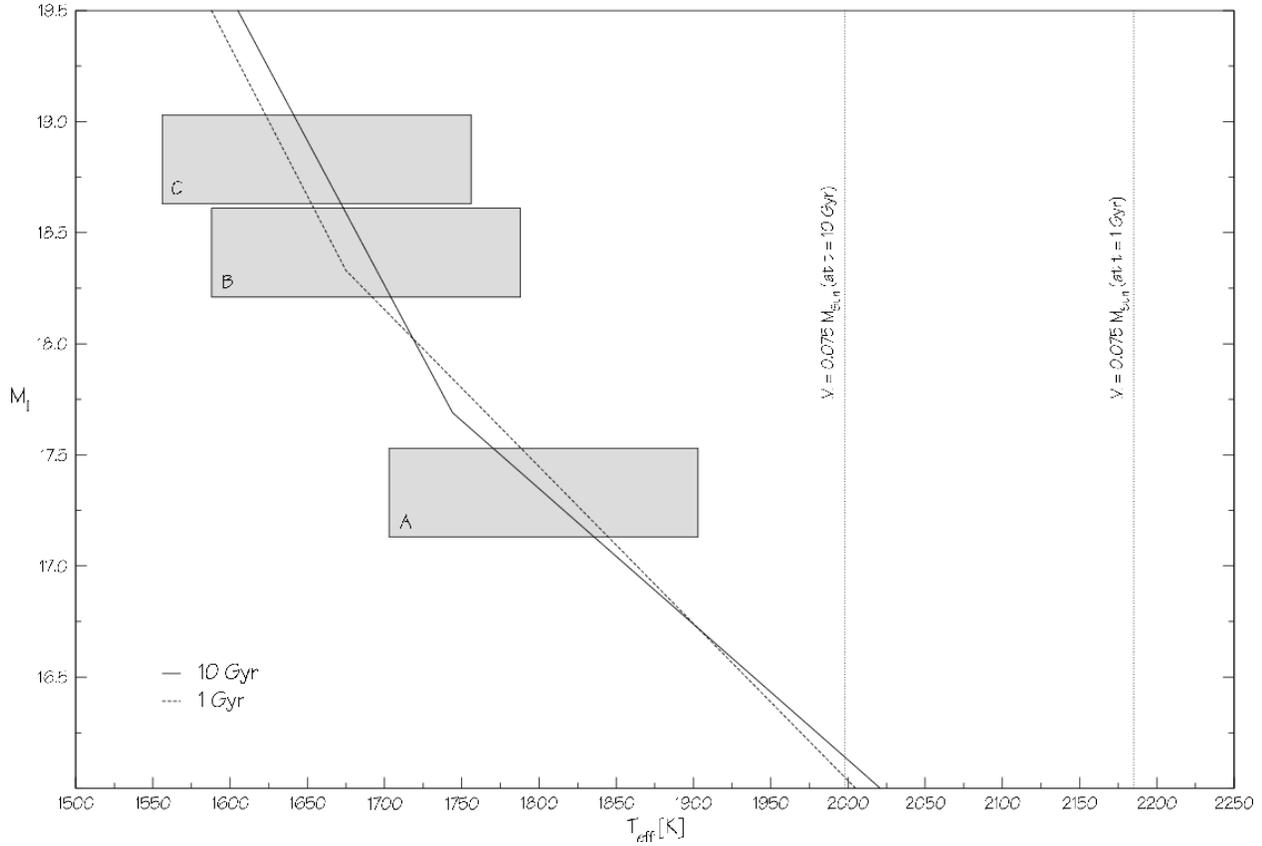}
      \caption{$M_{\mathrm{I_{C}}}$ vs Effective temperature. This figure shows the M$_{\mathrm{I_{C}}}$ as a function of the effective temperature, as given by the DUSTY models. Two isochrones (1 and 10~Gyr) are shown, together with the corresponding limits for sustained hydrogen burning. The position of the three components is represented by grey boxes, assuming $\pm$100~K uncertainty on the effective temperatures of the three components of DENIS-P~J020529.0-115925, and including the uncertainty on the distance for the M$_{\mathrm{I_{C}}}$. This figure shows that all three components must be substellar.\label{teff}}
   \end{figure*}

According to the DUSTY models \citep{2000ApJ...542..464C}, and assuming an age between 1 and 10~Gyr, Figure \ref{teff} shows that the absolute M$_{\mathrm{I_{C}}}$ magnitude of DENIS-P~J020529.0-115925 A corresponds to an effective temperature between 1\,700--1\,900~K, while B and C range between 1\,550--1\,800~K. These temperatures are consistent with the value reported by  \citet{2000ApJ...538..363B} for the unresolved system (1700$\sim$1800~K), and show that all components appear to be clearly substellar. According to the DUSTY models, the stellar/substellar limit is indeed around 2000~K at 10~Gyr and 2180~K at 1~Gyr, therefore warmer than any of the components of DENIS-P~J020529.0-115925.

Finally, the proper motion of the object \citep[$\sim$438 mas yr$^{-1}$, ][]{2002AJ....124.1170D} and the presence of these strong residuals at six different epochs spread over three years allow us to rule out definitively the eventuality of a coincidence with some background object.

\subsection{Dynamical Properties}
The separation between the primary and the secondary changed from $\sim$0\farcs390 to 0\farcs270 between October 2000 and December 2003, while the separation between the second and the third component is contained between $\sim$0\farcs075$\le \delta_{BC} \le$0\farcs055 (see Table \ref{pa_sep}). As stated by \citet{1968AJ.....73..190H} and then \citet{1977A&A....58..145S} in their analytical study, triple systems with moderate eccentricity and equal mass components are stable for ratios between the semi-major axes of the outer ($a_{2}$) and the inner orbits ($a_{1}$) greater than $a_{2}/a_{1} \ge$3.2. Assuming that the orbits of DENIS-P~J020529.0-115925 components have moderate eccentricities, and considering that the 3 components must have similar masses as explained above, DENIS-P~J020529.0-115925 would fit above the stability criterion, with a ratio between 3.6$\le a_{2}/a_{1} \le$7.1. The presence of a third component at the positions reported here is therefore dynamically possible.

The estimate of the separation corresponds to a semi-major axis of $\sim$1.2 A.U ($\sim$0\farcs075 at 19.7~pc). Corrected for a statistical factor of 1.26 as explained in \citet{1992ApJ...396..178F}, it leads to a ``best-guess'' semi-major axis of 1.9~A.U. According to Kepler's Third Law \citep{1609QB41.K32.......} and assuming a total mass of $\sim$0.1~M$_{\sun}$, the corresponding period, is $\sim$8 years.

\begin{figure*}
   \centering
   \includegraphics[width=\textwidth]{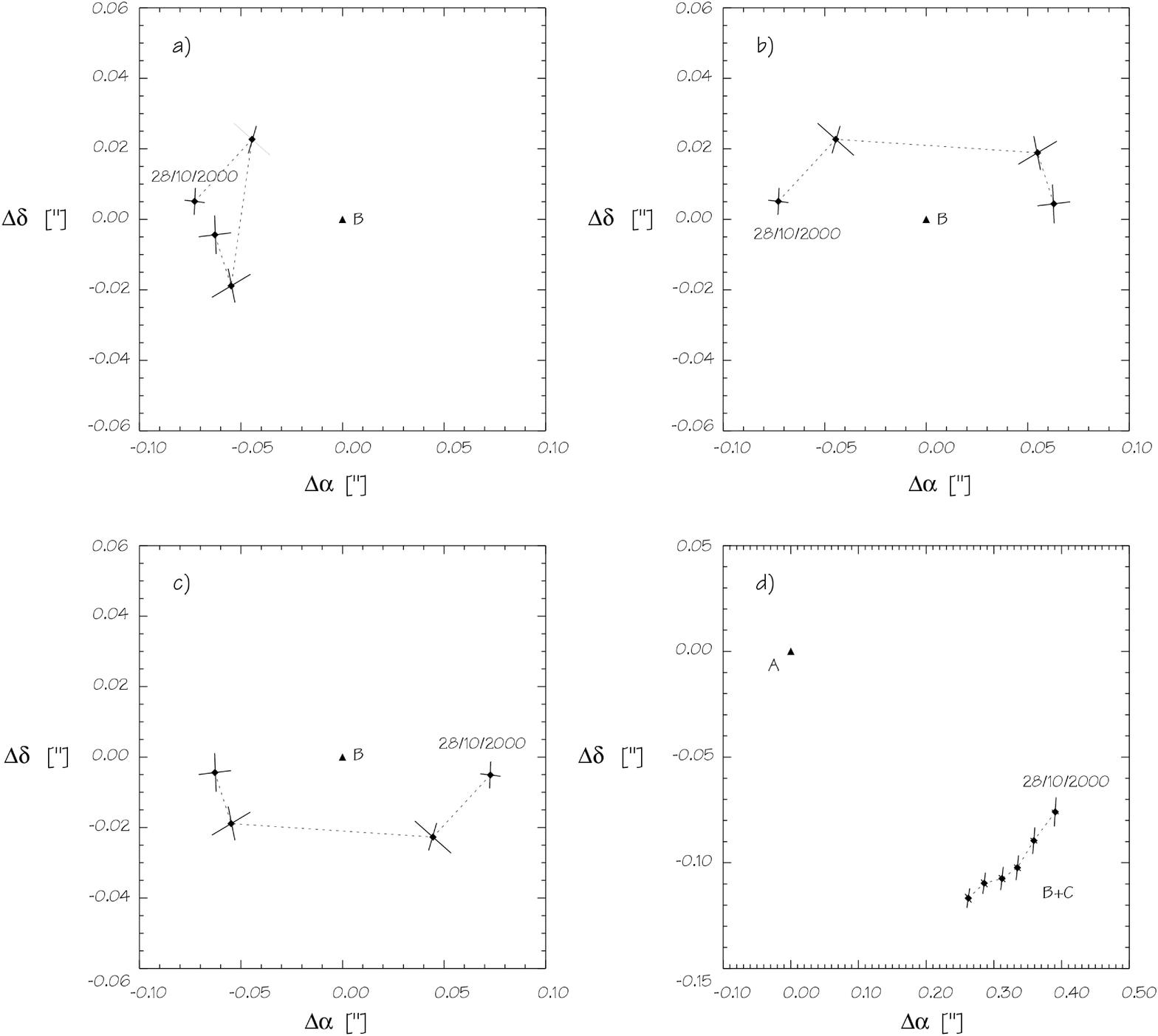}
      \caption{Panel a), b) and c): Relative motion of C (diamonds) around B (triangle). Panel d): Relative motion of B+C (diamonds) around A (triangle). 1-$\sigma$ uncertainties are indicated, as well as the first epoch. Epochs are linked by a dot line in chronological order. Due to the very short separation, there is an ambiguity of 180\degr\,  in the position angle of C around B. Panels a), b) and c) illustrate three configurations among the possible ones. a) raw results of the PSF fitting; b) epochs 1 and 2 from the raw results but 180\degr\, is added to epochs 3 and 4; c) epochs 3 and 4 from the raw results but 180\degr\, is added to epochs 1 and 2. While case a) cannot represent any Keplerian orbit, cases b) and c) could, with orbital periods comparable with the preliminary estimate we report here. Values from Table \ref{pa_sep}.}
      \label{motion}
   \end{figure*}

The dual-PSF fitting indicates that over the six epochs the separation between B and C is bounded between 0\farcs053--0\farcs074, while the position angle is bounded between  63\degr\, and 109\degr\, (see Table \ref{pa_sep}). The raw results of the PSF-fitting give a motion of C around B that cannot be orbital in nature (P.A does not vary monotically, as shown in panel a) of Figure \ref{motion}), but at such very short separations, the possible confusion between the primary and the secondary during the PSF-fitting yields to an ambiguity of $\pm$180\degr\, in the measurements of the position angle. Figure \ref{motion} illustrates two cases for which two epochs have been switched by 180\degr, giving possibly Keplerian orbital motions. The corresponding period (about half of the orbit in about 3~years) is in good agreement with that reported above (8~years).

With large and hardly assessable uncertainties and with 180\degr\, ambiguity on the P.A, we consider that the dual-PSF fitting is essentially a valuable sanity check showing that the elongation of the secondary is consistent with a binary. It gives a very rough and tentative estimate of the photometric and astrometric properties of the possible third component. New observations at higher angular resolution, using for example HST/ACS, ground based adaptive optics or speckle imaging, should be performed in order to confirm the multiplicity, and if confirmed to obtain more precise measurements of the motion of C around B.

\section{Conclusions}

We present here results of high angular resolution observations with HST/WFPC2 that allow us to conclude that DENIS-P~J020529.0-115925 is very likely to be a triple system of brown dwarfs. PSF subtraction on the secondary at six different epochs show unusual residuals in comparison with unresolved objects. Dual-PSF fitting shows that the shape of the secondary can be consistent with that of a binary, and that the motion of the possible third component could be Keplerian in nature. The configuration is consistent with a dynamically stable multiple system. 

Observations at higher angular resolution, using for example the HST/ACS, ground based adaptive optics or speckle imaging, should allow to confirm if DENIS-P~J020529.0-115925 is a triple system or not.

\acknowledgments
This work is based on observations collected with the NASA/ESA Hubble Space Telescope, obtained at the Space Telescope Science Institute, which is operated by the Association of Universities for Research in Astronomy, Inc., under NASA contract NAS 5-26555. These observations are associated with programs GO8720, GO9157, GO9345 and GO9968. This publication makes use of data from the DEep Near Infrared Survey. The DENIS project has been partly funded by the SCIENCE and the HCM plans of the European Commission under grants CT920791 and CT940627. It is supported by INSU, MEN and CNRS in France, by the State of Baden-Württemberg in Germany, by DGICYT in Spain, by CNR in Italy, by FFwFBWF in Austria, by FAPESP in Brazil, by OTKA grants F-4239 and F-013990 in Hungary, and by the ESO C\&EE grant A-04-046. This publication makes use of data products from the Two Micron All Sky Survey, which is a joint project of the University of Massachusetts and the Infrared Processing and Analysis Center/California Institute of Technology, funded by the National Aeronautics and Space Administration and the National Science Foundation. This work made use of data from the The Guide Star Catalogue~II, which is a joint project of the Space Telescope Science Institute and the Osservatorio Astronomico di Torino. 



Facilities: \facility{HST(WFPC2)}.




\clearpage






\clearpage

\begin{deluxetable}{lccc}
\tablecaption{Observation log.\label{obs_log}}
\tablewidth{0pt}
\tablehead{
\colhead{Filter} & \colhead{Exp. Time} & \colhead{Date Obs. [U.T]} & \colhead{Program} \\
\colhead{} & \colhead{[s]} & \colhead{\textsc{dd/mm/yyyy}} & \\
}
\startdata
	    F814W      & 600     & 28/10/2000  & GO8720 \\
	    F675W      & 300     & 28/10/2000  & GO8720 \\
	    F814W      & 1700    & 08/07/2001  & GO9157 \\
	    F814W      & 1800    & 21/01/2002  & GO9157 \\
	    F814W      & 400     & 14/07/2002  & GO9345 \\
	    F606W      & 1600    & 14/07/2002  & GO9345 \\
	    F814W      & 400     & 10/01/2003  & GO9345 \\	    
	    F814W      & 800     & 01/12/2003  & GO9968 \\
	    F606W      & 1000    & 01/12/2003  & GO9968 \\
\enddata
\end{deluxetable}

\clearpage

\renewcommand{\arraystretch}{1.5}
\begin{deluxetable}{lcc}
\tablecaption{Flux ratios and difference of magnitudes in the F814W filter\label{flux_ratio}}
\tablewidth{0pt}
\tablehead{
\colhead{Components} & \colhead{Flux Ratio} & \colhead{$\Delta$Mag}
}
\startdata
B/A   &   0.37       &  1.08         \\
C/A   &   0.25       &  1.50         \\
C/B   &   0.66       &  0.45         \\
\enddata
\tablecomments{Abs. Mag(A)=17.3~mag\\
The measurements reported here have large uncertainties, and should be considered with caution. The relative error from one image to the other is about 0.15~mag, but the real uncertainties can be higher. Measurements obtained on the GO8720 image.}
\end{deluxetable}

\clearpage

\begin{deluxetable}{lcc}
\tablecaption{Relative Astrometry of the three components\label{pa_sep}}
\tablewidth{0pt}
\tablehead{
\colhead{Date Obs.} & \colhead{Separation} & \colhead{P.A}
}
\startdata
\cutinhead{Component B with respect to A}
01/12/2003  &  0\farcs287$\pm$0\farcs005      &  246\degr$\pm$1\degr      \\
01/10/2003  &  0\farcs306$\pm$0\farcs005      &  249\degr$\pm$1\degr      \\
14/07/2002  &  0\farcs330$\pm$0\farcs005      &  251\degr$\pm$1\degr      \\
21/01/2002  &  0\farcs350$\pm$0\farcs005      &  253\degr$\pm$1\degr      \\
08/07/2001  &  0\farcs370$\pm$0\farcs005      &  256\degr$\pm$1\degr      \\
28/10/2000  &  0\farcs398$\pm$0\farcs005      &  259\degr$\pm$1\degr      \\
\cutinhead{Component C with respect to B}
28/10/2000  &  0\farcs073$\pm$0\farcs005      &  86\degr$\pm$3\degr      \\
08/07/2001  &  0\farcs050$\pm$0\farcs010      &  63\degr$\pm$5\degr      \\
21/01/2002\tablenotemark{1}  &  \nodata  &  \nodata     \\
14/07/2002  &  0\farcs058$\pm$0\farcs010      &  109\degr$\pm$5\degr      \\
10/01/2003  &  0\farcs063$\pm$0\farcs008       &  94\degr$\pm$5\degr      \\
01/12/2003\tablenotemark{1}  &  \nodata  &  \nodata     \\
\enddata
\tablecomments{Component C is barely resolved. The measurements reported for component C and their uncertainties are close to the limit of resolution of HST/WFPC2, and should be considered with caution.}
\tablenotetext{1}{The secondary is elongated, and the residuals after PSF subtraction are strong, but the PSF fitting does not give good enough results on the astrometry}
\end{deluxetable}

\clearpage


\clearpage

\clearpage

\clearpage

\clearpage

\clearpage

\end{document}